\documentclass[12pt]{article}
\usepackage{times}
\usepackage{amsmath}
\usepackage{amssymb}
\usepackage[pdftex]{graphicx} 
\usepackage{epstopdf}         
\usepackage{ITC}

\usepackage{ifpdf}
\ifpdf
\usepackage[pdftex]{graphicx} 
\usepackage{epstopdf}         
\else
\usepackage{graphicx}         
\fi

\usepackage{epsfig, amsmath,amssymb,epsf,cite,subfigure,scalefnt,epstopdf,setspace,fancyhdr,enumerate,mathrsfs}
\usepackage{xcolor}

\allowdisplaybreaks[4]

\usepackage{algorithm}
\usepackage{algorithmicx}
\usepackage{algpseudocode}

\title{\Large \vspace{-0.5in}
\MakeUppercase{Decorrelation Deep Learning for Fingerprint-Based Indoor Localization}}
\author{  Guojun~Xiong \\
    \normalsize  Department of Electrical Engineering and Computer Science  \\
    \normalsize The University of Kansas\\
    \normalsize Lawrence, KS£¬ 66045\\
    \normalsize gjxiong@ku.edu \\
     Faculty Advisors: \\
     Dr. Taejoon Kim and  Dr. Erik Perrins}

\date{}

\begin{document}

\maketitle 

\section{\MakeUppercase{Abstract}}

\noindent
Indoor localization is of particular interest due to its immense practical applications.
However, the rich multipath and high penetration loss of indoor wireless signal propagation make this task arduous.
Though recently studied fingerprint-based techniques can handle the multipath effects, the sensitivity of the localization performance to channel fluctuation is a drawback.
To address the latter challenge, we adopt an artificial multi-layer neural network (MNN) to learn the complex channel impulse responses (CIRs) as fingerprint measurements.
However, the performance of the location classification using MNN critically depends on the correlation among the training data.
Therefore, we design two different decorrelation filters that preprocess the training data for discriminative learning.
The first one is a linear whitening filter combined with the principal component analysis (PCA), which forces the covariance matrix of different feature dimensions to be identity.
The other filter is a nonlinear quantizer that is optimized to minimize the distortion incurred by the quantization.
Numerical results using indoor channel models illustrate the significant improvement of the proposed decorrelation MNN (DMNN) compared to other benchmarks.

\section{\MakeUppercase{Introduction}}
\noindent
The booming manufacture of smartphones and wearable devices has spurred the development of advanced localization techniques.
Better services to the users of smart devices can be offered if precise location information is available to the service providers \cite{yassin2016survey}.
There have been numerous indoor localization techniques that exploit surrounding networks (e.g., a group of base stations, access points, and local sensors) to calculate the coordinates of active devices by measuring uplink signals \cite{yassin2016survey, GSM_indoor, Fingerprint_Massive_MIMO}.
However, these ranging techniques pose critical challenges in an indoor environment due to the severe signal attenuation and rich multipath of indoor signal propagation\cite{Survey_localization}, limiting their accuracy.
Instead of ranging coordinates, classifying a location zone in an indoor context (e.g., room or office numbers) was the focus of the work in \cite{K-nearest,GSM_indoor,Fingerprint_Massive_MIMO}, referred to as  fingerprint-based localization technique.

~\\
\noindent
The fingerprint-based techniques adopt a radio map by matching an online measurement to the offline training measurements  \cite{K-nearest, GSM_indoor,Fingerprint_Massive_MIMO}.
The K-nearest neighbors \cite{K-nearest} and support vector machine (SVM) \cite{GSM_indoor} have been popularly studied.
Though these approaches are immune to multipath effects, they have inherited limitations \cite{yassin2016survey, Survey_localization}.
First, the localization performance considerably varies with the choice of fingerprint measurements, such as time-of-arrival (ToA) and received signal strength (RSS), at different locations.
Second, the performance is sensitive to channel fluctuations.
Third, the ToA or RSS fingerprint measurements are susceptible to in-band interference.
These limitations constitute obstacles in the practical application of fingerprint-based localization techniques.

~\\
\noindent
In this paper, we explore an artificial multi-layer neural network (MNN)-based machine learning framework to overcome the above limitations.
We go beyond the prior approaches\cite{K-nearest, GSM_indoor, Fingerprint_Massive_MIMO}, and propose a framework that learns the channel impulse responses (CIRs), which are measured by multiple distributed sensors.
Because the CIR captures the particulars of the channel propagation environment, it can be exploited to enhance the indoor localization performance \cite{yassin2016survey}.
Main reasons for the limited application of MNN to localization are the large training overhead and the restrictive degree of classification accuracy, which is caused by the high correlation between channel measurements collected from adjacent regions.
If the training dataset is highly correlated, a deep neural representation after training is also highly entangled across nodes because any subtle fluctuations in the correlated input will modify most of the entries in the MNN, leading to unreliable classification performance \cite{Bengio12}.
To tackle this issue, we propose, in this paper, a class of decorrelation filters applied to the training dataset prior to training MNN.

~\\
\noindent
We design two different types of decorrelation filters.
The first one is whitening transformation combined with principal component analysis (PCA), which is linear and invertible.
The PCA algorithm finds the major features of the training data and the whitening filter removes the correlation among the features.
The second decorrelation filter employs a nonlinear quantization transformation.
By leveraging statistical information of the training data, we  design an optimal quantizer that minimizes the distortion caused by the quantization.
We numerically show that the proposed decorrelation MNN (DMNN) remarkably outperforms both the conventional MNN and SVM \cite{GSM_indoor} and it achieves an inspiring mis-classification rate around  $3-4\%$.


\subsubsection*{Notation}
\noindent
A bold lower case letter $\bold{a}$ is a vector, a bold capital letter $\bold{A}$ is a matrix, and $\bold{I}_M$ denotes the $M\times M$ identity matrix.
$\bold{a}(i:j)$ denotes the subvector formed by taking the $i$th to $j$th entries of $\bold{a}$, and $\bold{A}(i, j)$ is the $i$th row $j$th column element of $\bold{A}$.
$\bold{A}^T$ represents the transpose, $\text{Vec}[\mathcal{A}]$ stacks all the elements in the set $\mathcal{A}$ into a column vector, and $\mathrm{Card}(\mathcal{A})$ is the cardinality of $\mathcal{A}$.
$\text{Var}(\bold{a})$ is the covariance matrix of $\bold{a}$, and $\mathbb{E}(\cdot)$ is the expectation. We denote $\Re(a)$ and $\Im(a)$ as the real part and imaginary part of a complex number $a$, respectively.
Given $f(\cdot): \mathbb{R} \mapsto \mathbb{R}$, we denote for a vector $\bold{a}\in\mathbb{R}^{n\times 1}$, $f(\bold{a}) = [f(\bold{a}(1)), \ldots, f(\bold{a}(n)) ]^T\in \mathbb{R}^{n\times 1}$.


\section{\MakeUppercase{System Model and Problem Formulation}} \label{Section2}

\subsection*{\MakeUppercase{System Model}}

\begin{figure}[h]
\vspace{-0.5cm}
\centering
\includegraphics[width=9cm, height=4.0cm]{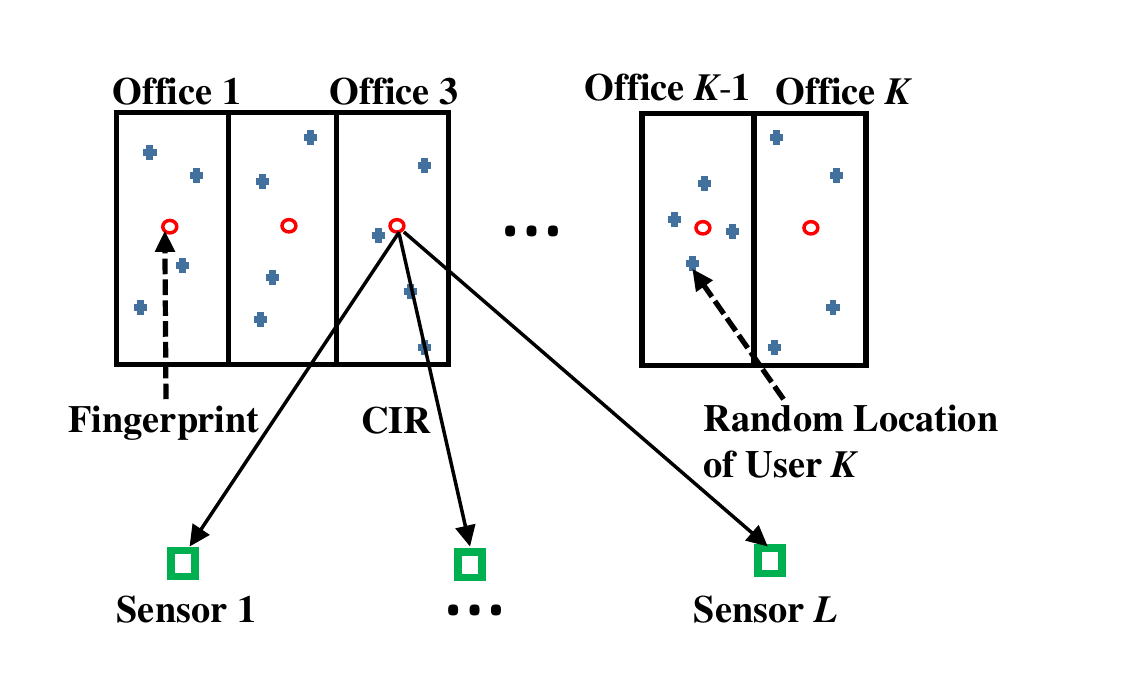}
\vspace{-0.2cm}
\caption{System model of an indoor office environment. } \label{Office_model}
\end{figure}

\noindent
We consider an indoor uplink channel from a single-antenna user to $L$ distributed single-antenna sensors as shown in Figure~\ref{Office_model}.
There are $K$ clustered and equal-sized offices in Fig.~\ref{Office_model}.
Each user is located in an office and the sensors are distributed uniformly.
Define the fingerprints as the CIRs measured from the center of each office room to the distributed sensors.
We model the CIR measurement from the $k$th fingerprint location to the $\ell$th distributed sensor as a wideband multipath block-fading channel \cite{RSStoCIR},
\begin{equation}\label{CIR}
\bold{h}_{k\ell}(t)=[h_{k\ell}^1(t), h_{k\ell}^2(t), \ldots, h_{k\ell}^D(t)],
\end{equation}
where $t$ denotes the $t$th time slot, $D$ is the number of multipaths, and $k=1, \ldots, K$ and $\ell=1, \ldots, L$.
The $h_{k\ell}^d(t)$ in \eqref{CIR} can be rewritten as
$
{h_k}_{\ell}^d(t)=A_{k\ell}^d(t)e^{j\theta_{k\ell}^d(t)},~ d\in\{1, 2,\ldots, D\},
$
where $A_{k\ell}^d(t)=|h_{k\ell}^d(t)|$ and $\theta_{k\ell}^d(t)=\mathrm{arg}(h_{k\ell}^d(t))$ with $\theta_{k\ell}^d(t)$ being a uniform random variable, uniformly distributed  in the interval $[0, 2\pi]$.
We further model $A_{k\ell}^d(t)$ as \cite{Fingerprint_Massive_MIMO}
\begin{equation}\label{CIR_model}
A_{k\ell}^d(t)=h_{k\ell}^{d, \text{small}}\sqrt{h_{k\ell}^{d, \text{large}}(t)}.
\end{equation}
$h_{k\ell}^{d,\text{large}}(t)\in \mathbb{R}$ is the large-scale fading component and it follows the multi-wall room-level channel propagation model in $3.5$ GHz \cite{Fading_channel_model, office_channel_model}, which is given by in dB,
\begin{equation}\label{large-scale-fading}
h_{k\ell}^{d,\text{large}}(t)=P_s^d(t)-(2\text{log}_{10}r+3n_{\text{wall}}+0.2n_{\text{door}})~[dB],
\end{equation}
where $P_s^d(t)$ [dB] is the  effective isotropic radiated power for the $d$th path, $r$ [m] is the distance from the indoor user to the sensor in meters,
the pre-log factor $2$ in \eqref{large-scale-fading} denotes the $2$ dB loss per meter in free space, $n_{\text{wall}}$ is the number of concrete walls ($3$ dB loss per wall), and $n_{\text{door}}$ is the number of doors ($0.2$ dB loss per door).
We assume in this work $P_s^d(t)$ follows the international telecommunication union (ITU) indoor office channel model in \cite{ITUchannel}.
$h_{k\ell}^{d, \text{small}}$ is a stationary, independent, and identically distributed Gaussian random variable, i.e, $h_{jk}^{d, \text{small}} \sim \mathcal{N}(\mu,\sigma^2),$ modeling indoor small-scale fading caused by the user's random position within an office.
Because the CIRs in adjacent offices experience quite similar signal propagation environments, the amplitudes of them are highly correlated.

~\\
\noindent
The CIR estimation  is a well-studied area with various  accurate and practical estimation algorithms being developed \cite{Kim, LTE_book}.
Hence, we assume, in this work, perfect CIR knowledge available and focus on the development of a fingerprint-based localization technique.

\subsection*{ \MakeUppercase{Supervised Learning for Localization}}
\noindent
The goal is to directly classify the office number in Figure \ref{Office_model} by observing uplink CIRs from a wireless device in the office.
This regression problem can be formulated as a supervised machine learning problem that learns a specific office number from the offline measurements and then takes the online measurement to classify the office number.
~\\

\noindent
We denote the offline training data set as ${\mathcal{D}}=\{{\mathcal{D}}_1, \ldots, {\mathcal{D}}_K\}$ where ${\mathcal{D}}_k=\{(\bold{x}_k(t), \bold{y}_k(t))\}_{t=1}^N$ and $N$ is the number of  input-output training data pairs.
The input in ${\mathcal{D}}_k$ is given by
\begin{equation} \nonumber
\bold{x}_k(t)=\text{Vec}[\{ ( \Re(h_{k\ell}^d(t)), \Im(h_{k\ell}^d(t)))  |d=1,\ldots, D, \ell=1, \ldots, L\}] \in \mathbb{R}^{2DL \times 1}, \forall t.
\end{equation}
The output $\bold{y}_k(t)$ in $\mathcal{D}_k$ is given by $\bold{y}_k(t)=\bold{e}_k$, where $\bold{e}_k$ is the $k$th column of $\bold{I}_K$.
Then, the supervised learning task is equivalent to finding a mapping function $f$, which maps the $2DL\times 1$ dimensional input data $\bold{x}_k(t)$ to $\bold{y}_k(t)$,
$$ f(\cdot): \bold{x}_k(t)\mapsto \bold{y}_k(t)\in S_\bold{y}=\{\bold{e}_1, \ldots, \bold{e}_K\},  ~\forall k, t. $$
Based on the training data set $\mathcal{D}$, the $f(\cdot)$ design problem can be formulated as
\begin{equation}\label{MMSE}
\tilde{f}(\cdot)=\underset{{f}(\cdot)}{\text{argmin}}~~ \frac{1}{NK}\sum\limits_{k=1}^{K}\sum\limits_{t=1}^{N}\|f(\bold{x}_k(t))-\bold{y}_k(t) \|_2^2.
\end{equation}
There exist numerous approaches that  handle the problem in \eqref{MMSE}.
 Among those, artificial MNNs \cite{Goodfellow-et-al-2016} has recently received tremendous attention due to its capability to learn  complex relationships between input $\bold{x}_k(t)$ and output $\bold{y}_k(t), ~\forall k, t.$

\begin{figure}[h]
	\setlength{\abovecaptionskip}{-2mm}
	\centering
	\includegraphics[width=8.5cm, height=4.8cm]{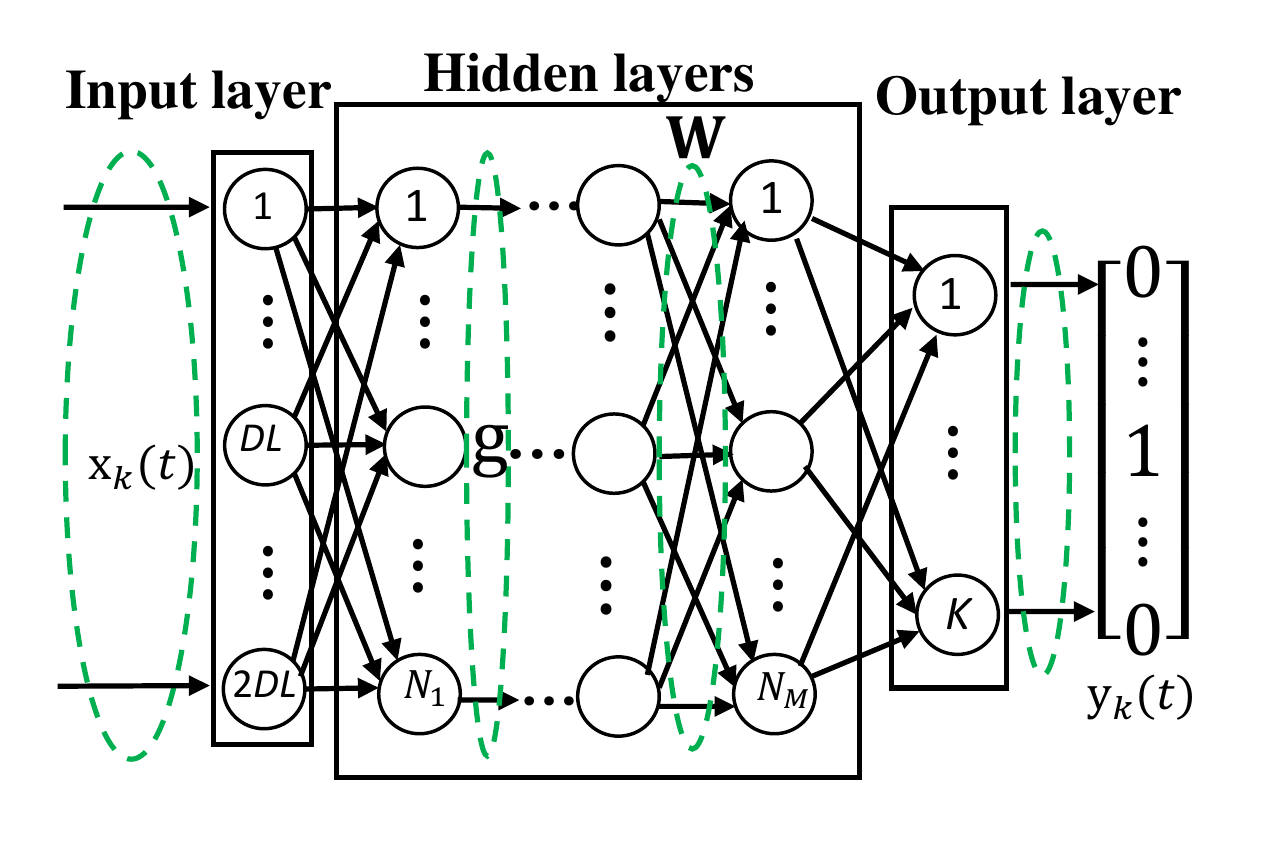}
	\caption{Block diagram showing a  multi-layer neural network (MNN).    } \label{FNN}
\end{figure}

\noindent
MNN is a deep learning model that consists of three parts, i.e., the input layer, hidden layer, and  output layer as shown in Figure~\ref{FNN}.
The numbers of neurons at the input layer and the output layer are determined by the dimensions of the input data ($\bold{x}_k(t) \in \mathbb{R}^{2DL \times 1}$) and the output data ($\bold{y}_k(t) \in \mathbb{R}^{K\times 1}$), respectively.
 It models the mapping function $f(\cdot)$ via multiple nonlinear transformations of $M$ hidden layers.
The $m$th hidden layer represents a mapping $g_m(\bold{r}_{m-1}): \mathbb{R}^{N_{m-1}}\mapsto \mathbb{R}^{N_m}$, where $\bold{r}_{m-1}\in\mathbb{R}^{N_{m-1}}$ is the output of the $(m-1)$th hidden layer, and  the mapping function is given by
 \begin{equation}\label{full_connect}
 \bold{r}_m=g_m(\bold{r}_{m-1})=\sigma(\bold{W}_m\bold{r}_{m-1}+\bold{b}_m).
 \end{equation}
In \eqref{full_connect}, $\bold{W}_m\in\mathbb{R}^{N_m\times N_{m-1}}$ is the weight matrix and $\bold{b}_m\in\mathbb{R}^{N_m}$ is the bias vector at the $m$th hidden layer. The $\sigma(\cdot)$ in \eqref{full_connect} is an activation (simple) function.

~\\
\noindent
There are several choices for activation function $\sigma(\cdot)$, such as the linear $\sigma(x)=x$, sigmoidal $\sigma(x)=\frac{1}{1+e^x}$, and rectified linear unit (ReLU)  $\sigma(x)= \text{max}(0, x)$.
In general, the linear activation function is applied to the input layer, the sigmoidal function is used for predicting the probability, and the ReLU function is used to increase the sparsity of the weight matrices\cite{NN_book_classfication}.
For multiclass learning and classification, the sigmoidal activation function has been proven to be effective \cite{NN_book_classfication}.
The stochastic gradient descent (SGD) algorithm \cite{Goodfellow-et-al-2016} is a popularly-adopted MNN training approach that
designs the set of weight matrices and bias vectors $\{\bold{W}_m, \bold{b}_m\}_{m=1}^M$ of MNN in Figure \ref{FNN}, based on the training data set $\mathcal{D}$.

\subsection*{\MakeUppercase{General Description of The Proposed Technique}}
\begin{figure}[h]
\vspace{-0.5cm}
\setlength{\abovecaptionskip}{-2mm}
\centering
\includegraphics[width=6.5cm, height=3.3cm]{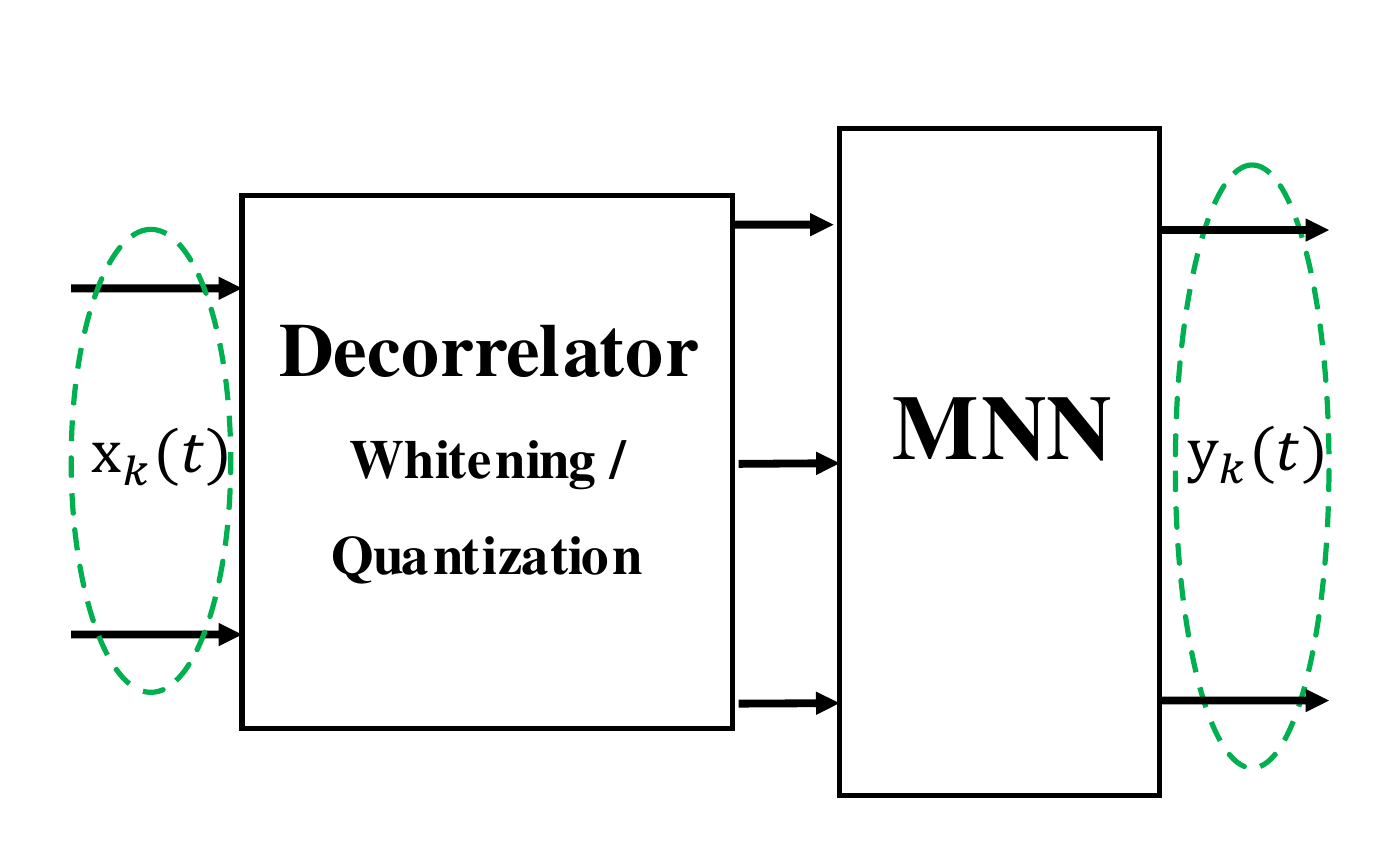}
\caption{Block diagram showing the proposed decorrelation MNN (DMNN).    } \label{DFNN}
\vspace{-0.5cm}
\end{figure}

\noindent
Though the traditional MNNs can be employed directly for our indoor localization task, the collected CIRs from one office area is highly correlated with those from adjacent office areas  because of richly scattered indoor wave propagation environments.
When an MNN is trained in such an environment, it reders a highly entangled representation for the trained MNN, which critical deteriorates the classification performance \cite{Bengio12}.
The proposed remedy is to design a decorrelation filter that reduces the correlation among the training set $\mathcal{D}$ prior to training.
The entire proposed DMNN structure is illustrated in Figure \ref{DFNN}.

\indent
%
%
\section{\MakeUppercase{Decorrelation MNN for indoor Localization}} \label{Section3}
\noindent
In this section, we propose a DMNN model for indoor localization as shown in Figure~\ref{DFNN}.
Two decorrelation filters to preprocess the input training data are proposed, one by using the PCA-Whitening technique \cite{Whitening}, which is linear, and the other by leveraging the optimal quantizer design \cite{Quantization_pages}, which is non-linear.

\subsection*{\MakeUppercase{Linear Decorrelation Filter Design}}
\noindent
A whitening filter is a linear projection that decorrelates each element of a random vector by transforming its covariance matrix into the identity matrix.
We let the training input data $\{\{\bold{x}_k(t)\}_{t=1}^{N}, \forall k \}$ be $KN$ realizations of a random vector $\bold{x}=[ x_1, \ldots, x_{2DL}]^T$ whose covariance matrix is given by Var$(\bold{x})=\pmb{\Sigma}\in \mathbb{R}^{2DL\times 2DL}$.
Then a whitening filter $\bold{T}\in \mathbb{R}^{2DL\times 2DL}$ is designed to transform $\bold{x}$ to a random vector $\bold{z}=\bold{T}\bold{x}$
such that $\text{Var}(\bold{z})=\bold{I}_{2DL}$, which is satisfied if the whitening matrix $\bold{T}$ obeys
\begin{equation}\label{requirement_WMatrix}
\bold{T}\pmb{\Sigma}\bold{T}^T=\bold{I}_{2DL}.
\end{equation}
Note that the $\bold{T}$ satisfying \eqref{requirement_WMatrix} is not unique  because any left rotation of $\bold{T}$,  $\bold{R}\bold{T}$, where $\bold{R} \in \mathbb{R}^{2DL\times 2DL}$ is a unitary matrix,  also satisfies \eqref{requirement_WMatrix}.
Hence, a general form of the whitening matrix is expressed as $\bold{T}=\bold{R}\pmb{\Sigma}^{-\frac{1}{2}}$, where $\bold{R}\bold{R}^T=\bold{R}^T\bold{R}=\bold{I}_{2DL}$.

~\\
\noindent
An important criterion when desining a whitening filter is to maintain the variance of $\bold{z} = \bold{T}\bold{x}$ to be as close as the variance of $\bold{x}$ \cite{optimal_whitening}.
For instance, if the values of $\bold{x}$ increase, the values of $\bold{z}$ vary proportionally to $\bold{x}$.
The similarity between $\bold{z}$ and $\bold{x}$ can be reasonably quantified by the cross-covariance matrix
$$\pmb{\Phi}\triangleq\text{cov}(\bold{z},\bold{x})=\bold{T}\pmb{\Sigma}=\bold{R}\pmb{\Sigma}^{\frac{1}{2}}\in\mathbb{R}^{2DL\times 2DL}.$$
Hence, the goal is to optimize $\bold{R}$ to maximize the minimum entry of $\pmb{\Phi}$, i.e.,
\begin{equation}\label{diag_maximization}
\bold{R}^{o}={\text{argmax}}_{\bold{R}}~ \text {min}_{i, j}~ \pmb{\Phi}(i, j)~~~\text{subject to}~~ \bold{R}^T\bold{R}=\bold{R}\bold{R}^T=\bold{I}_{2DL}.
\end{equation}
However, the problem in \eqref{diag_maximization} does not admit a closed-form expression.
To circumvent, we propose to relax the criterion in \eqref{diag_maximization} to maximizing the sum of squared elements of $\pmb{\Phi}$, i.e.,  $\sum\limits_{i=1}^{2DL}\sum\limits_{j=1}^{2DL}\pmb{\Phi}(i,j)^2=\sum\limits_{i=1}^{2DL}\sum\limits_{j=1}^{2DL}\text{cov}(z_i,x_j)^2. $
Hence, it turns to the squared frobenius norm maximization problem:
\begin{equation}\label{Fnorm_max}
\bold{R}^{\star}={\text{argmax}}_{\bold{R}}~ \|\pmb{\Phi}\|_F^2={\text{argmax}}_{\bold{R}} ~\mathrm{tr}(\bold{R}\pmb{\Sigma}\bold{R}^T)
~~~\text{subject to}~~ \bold{R}^T\bold{R}=\bold{R}\bold{R}^T=\bold{I}_{2DL}.
\end{equation}
Given the singular value decomposition $\pmb{\Sigma}= \bold{U}\pmb{\Lambda}\bold{U}^T$, where
$\bold{U}^T\bold{U}=\bold{U}\bold{U}^T=\bold{I}_{2DL}$ and $\pmb{\Lambda}\in \mathbb{R}^{2DL \times 2DL}$ is a singular value  matrix with the $i$th largest singular value at position $(i, i)$ for $i=1, \ldots, 2DL$, the solution to \eqref{Fnorm_max} is
$\bold{R}^{\star}=\bold{U}^T,$
which gives the whitening filter
\begin{equation}
\bold{T}=\pmb{\Lambda}^{-\frac 1 2}\bold{U}^T.
\end{equation}

\subsection*{\MakeUppercase{Nonlinear Decorrelation Filter Design}}
\noindent
An alternative method for decorrelating the training data is quantization.
A well-designed quantizer can force highly correlated data into a distinct discrete grid to make them separable  \cite{Quantization_pages}.
We assume for tractability that the elements of the training input data $\{\{\bold{x}_k(t)\}_{t=1}^{N}, \forall k \}$ are $2KDLN$ realizations of a random variable $x$ that follows Gaussian distribution with mean $\upsilon$ and variance $\varsigma^2$, i.e., $x\sim \mathcal{N}(\upsilon, \varsigma^2)$.
Here, the Gaussian assumption simplifies the analysis and quantization algorithm development.
Define $\Psi=\{q_1, \ldots, q_M\}$ as a scalar quantization grid with level $M$,
and then the real line $\mathbb{R}$ is partitioned into $M$ disjoint intervals as $\{\mathcal{V}_i\}_{i=1}^{M}$ where
\begin{equation}
\mathcal{V}_i=\{x\in\mathbb{R}: |x-q_i|={\text{min}}_{1\leq j\leq M}|x-q_j|\}.
\end{equation}
Mathematically, the element-wise quantization function $Q(x): x\in\mathbb{R}\mapsto \Psi$ with $M$ distinct levels can be expressed as
$Q(x)\triangleq \sum\limits_{i=1}^{M} q_i\pmb{1}_{\mathcal{V}_i},$
where $\pmb{1}_{\mathcal{V}_i}$ is an indicator function.

~\\
\noindent
The resulting squared quantization distortion of $Q$, i.e., the squared difference between the input and the quantized value, is  given by
\begin{equation}\label{Quan_distortion}
e_Q(\Psi)=\mathbb{E}|x-Q(x)|^2
=\sum\limits_{i=1}^{M}\int_{\mathcal{V}_i}\!\!\frac{|q_i-\tau|^2}{\sqrt{2\pi\varsigma^2}}e^{-(\tau-\upsilon)^2/2\varsigma^2}d\tau.
\end{equation}
Then an optimal quantizer design problem is to find a finite set $\Psi=\{q_1, \ldots, q_M\}$ by minimizing the  distortion in \eqref{Quan_distortion}, namely,
\begin{equation}
\Psi^{\star}={\text{argmin}_{\Psi}}~ e_Q(\Psi)~~~\text{subject to}~~ \mathrm{Card}(\Psi)=M.
\end{equation}

~\\
\noindent
Since the distortion function in \eqref{Quan_distortion} is continuously twice differentiable,  an optimal quantizer is obtained by finding $\Psi$ that satisfies
\begin{equation}\label{stationary_quantizer}
\frac{\partial e_Q(\Psi)}{\partial q_i}=0,  ~ 1\leq i\leq M.
\end{equation}
The fixed point equations in \eqref{stationary_quantizer} can be numerically solved by using a line search algorithm.
In this paper, we adopt the Newton-Raphson method to iteratively solve them.
We let $[(q_i+q_{i-1})/2~ ~(q_i+q_{i+1})/2]$ be each partition of $\mathcal{V}_i$.
Then we have
\begin{equation}
e_Q(\Psi)=\sum\limits_{i=1}^{M}\int_{(q_i+q_{i-1})/2}^{(q_i+q_{i+1})/2}\frac{|q_i-\tau|^2}{\sqrt{2\pi\varsigma^2}}e^{-(\tau-\upsilon)^2/2\varsigma^2}d\tau.
\end{equation}

~\\
\noindent
Starting from the initial $\{q_1^{(0)}, \ldots, q_M^{(0)}\}$, the Newton-Raphson method iterates for $n=1,2,\ldots$ the recursion equation
\begin{equation}\label{Netwon_Raphson}
q_i^{(n+1)}=q_i^{(n)}-\alpha^{(n+1)}\left[\frac{\partial^2 e_Q(\Psi)}{\partial^2 q_i^{(n)}}\right]^{-1}\frac{\partial e_Q(\Psi)}{\partial q_i^{(n)}}, \forall i,
\end{equation}
where $\alpha^{(n+1)}$ is the step size.
It is shown that the recursion in \eqref{Netwon_Raphson} converges to the solution $\{q_1^{\star}, \ldots, q_M^{\star}\}$ based on the central limit theorem\cite{GaussianCase}.
Finally, after a sufficient amount of iterations, the designed quantizer $\Psi^{\star}=\{q_1^\star, \ldots, q_M^\star\}$ is used to quantize the input of MNN.

\section{\MakeUppercase{Numerical Simulations}}\label{Section4}
\begin{figure}[h]
\setlength{\abovecaptionskip}{-2mm}
\centering
\includegraphics[width=12.5cm, height=5.8cm]{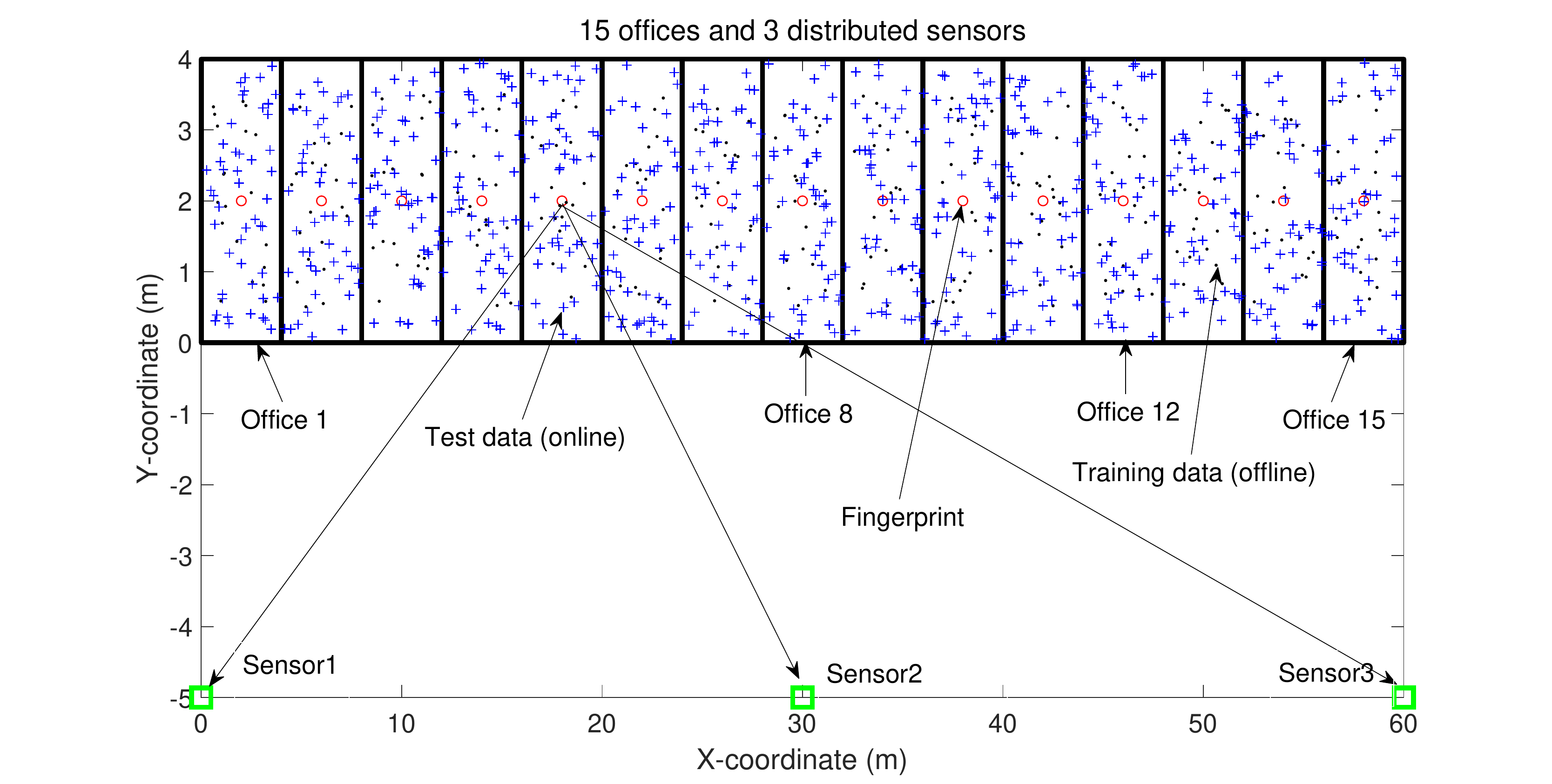}
\caption{Layout of an indoor office environment with 15 offices and  3 sensors.} \label{cell_model}
\end{figure}

\noindent
In this section, we numerically demonstrate the proposed DMNN techniques for indoor localization.
An indoor office  environment  with length $60$ m and width  $9$ m is modeled as shown in Figure \ref{cell_model}.
There are $K\!=\!15$ offices of equal size, $4$ m $\times$ $4$ m, and $L\!=\!3$ distributed sensors with  $30$ m separation.
It also shows  $15$  fingerprints ($\circ$) and  offline training measurements locations ($+$).
We set the mean and variance of $h_{k\ell}^{d, \text{small}}, \forall \ell,~ k,~ d$ as  $(\mu, \sigma^2)=(1, 0.1)$ and the number of multipath as $D=3$.
The MNN in Figure~\ref{DFNN} is implemented to have 4 hidden layers, where the numbers of neurons of the 4 hidden layers are, respectively, $100-200-100-50$.
Specifically, the first hidden layer is chosen as ReLU layer and others are sigmoidal layers.
We collect $N=100$ training measurements at the training signal-to-noise ratio (SNR) $20$ dB.
Based on the ITU indoor office channel model\cite{ITUchannel}, we set $P_s^1(t)=20$ dB, $P_s^2(t)=16$ dB, and $P_s^3(t)=10$ dB in \eqref{large-scale-fading}, $\forall t$.

\begin{figure}[h]
\vspace{-0.2cm}
\centering
\includegraphics[width=8cm, height=5.6cm]{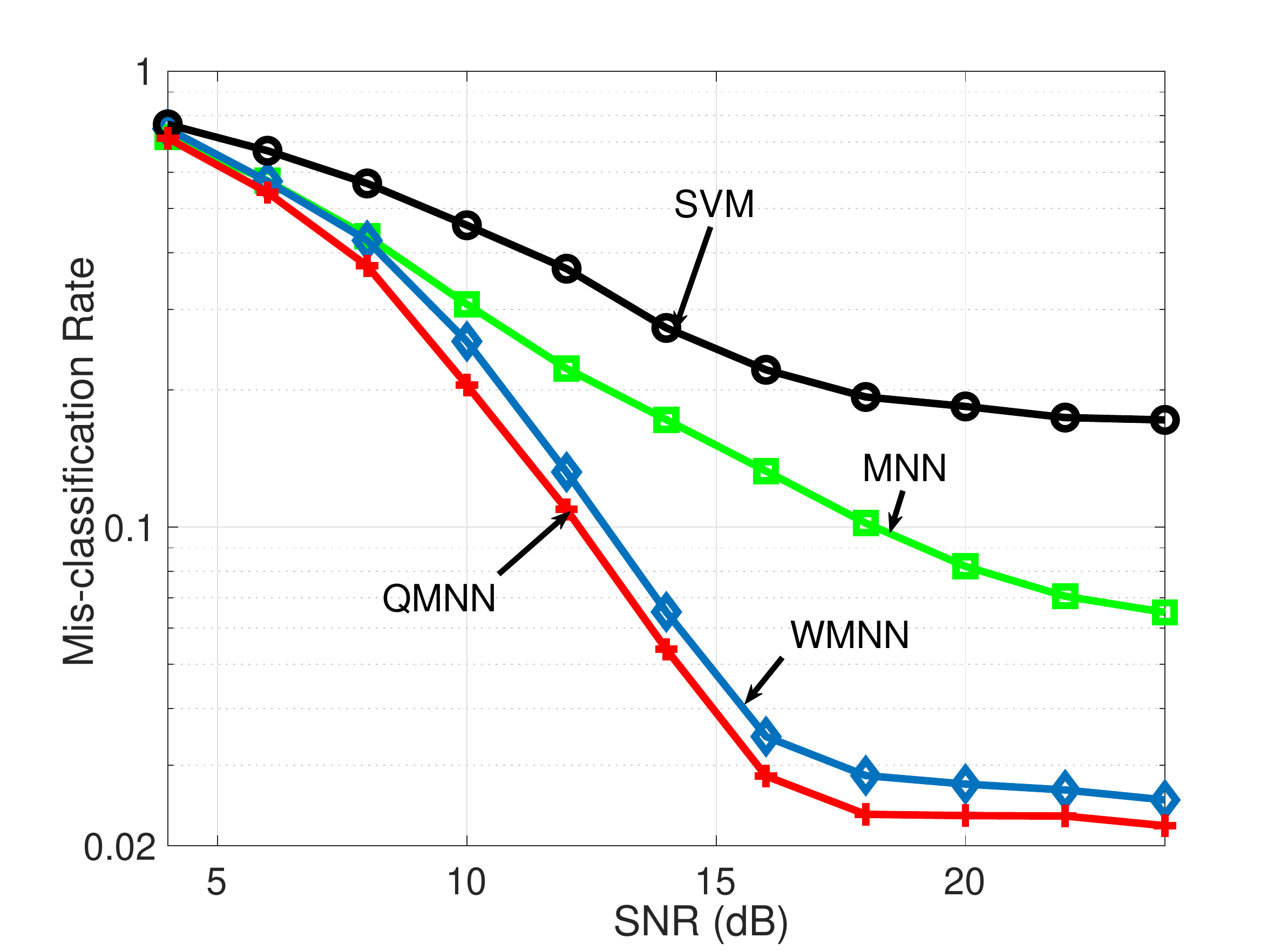}
\caption{Comparison of the proposed DMNNs with SVM and MNN $(K=15, L=3, D=3, N=100)$. } \label{Decorrelation_filter_performance}
\end{figure}

~\\
\noindent
Figure \ref{Decorrelation_filter_performance} compares the mis-classification rate of the proposed DMNN to MNN in Figure \ref{FNN} and SVM in \cite{GSM_indoor}, where MNN is the same as the proposed DMNN except for the decorrelation filter.
In Figure \ref{Decorrelation_filter_performance}, we denote DMNN with quantization as QMNN and DMNN with whitening as WMNN.
$M=10$ distinct quantization levels are considered for QMNN.
 Interestingly, WMNN and QMNN remarkably outperform both benchmarks MNN and SVM.
 QMNN and WMNN achieve a mis-classification rate of around $3$--$4 \%$ at SNR $=20$ dB, which demonstrates significant performance enhancement of the proposed DMNN.

\begin{figure}[h]
\centering
\includegraphics[width=8cm, height=5.6cm]{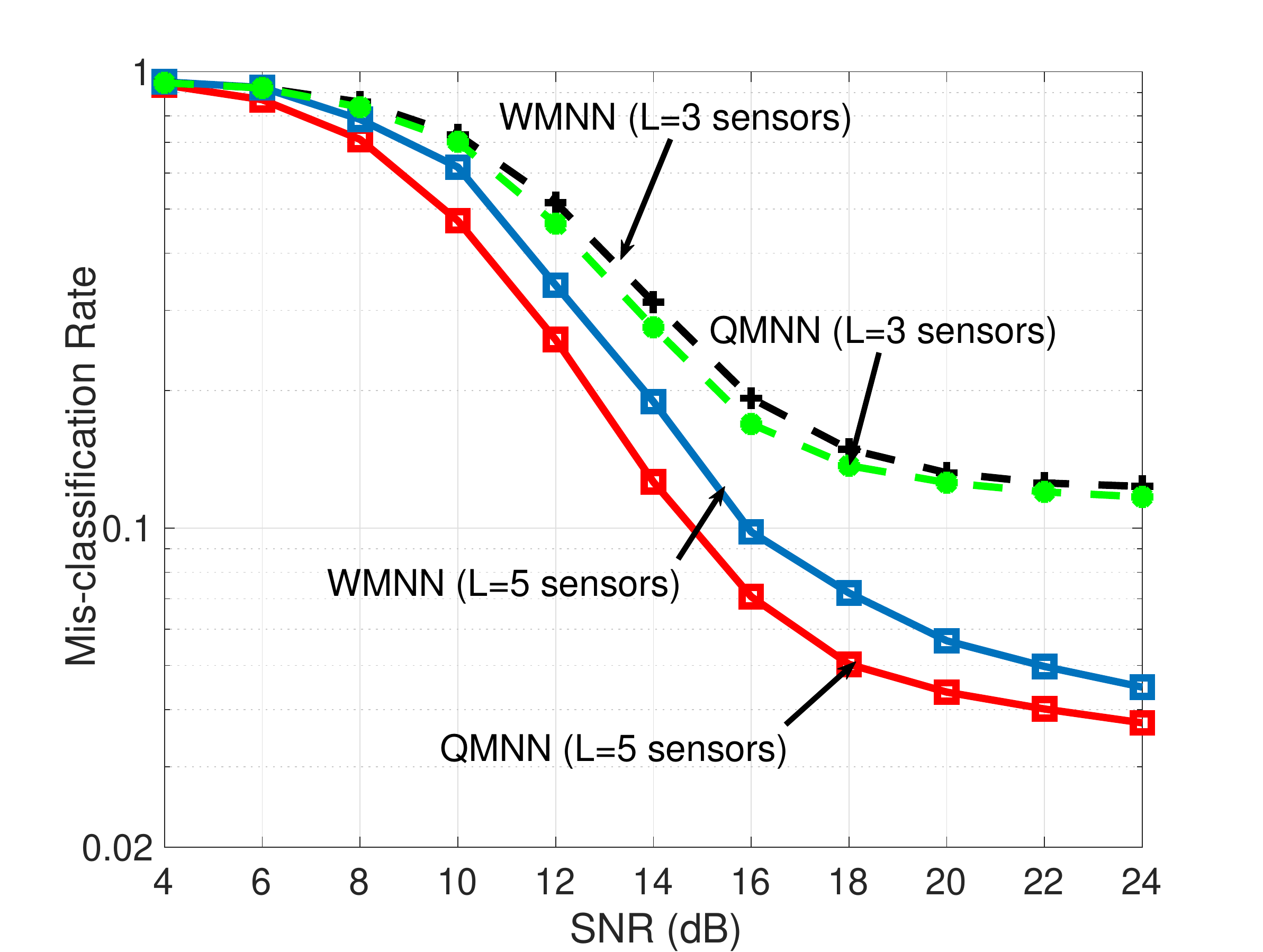}
\caption{The proposed DMNNs with different number of sensors $(K=30, D=3, N=100)$. } \label{Dense_system}
\end{figure}

~\\
\noindent
We also evaluate the impact of the  number of distributed sensors on the localization performance when the number of offices is doubled, i.e., $K=30$ in Figure \ref{Dense_system}.
It is shown in Figure \ref{Dense_system} that the proposed DMNNs can achieve a   mis-classification rate around $12\%$ when the number of distributed sensors are kept $L=3$ as in Figure \ref{Decorrelation_filter_performance}.
However, when the number of sensors increases to $L=5$, the proposed DMNNs achieve the  mis-classification rate around $3$--$4 \%$.
A more number of distributed sensors improve the classification performance because of the enhanced spatial diversity.

\section{\MakeUppercase{Conclusions}}\label{Section5}
\noindent
In this paper, we proposed DMNN-based indoor localization techniques to address the previously beset indoor localization challenges.
To reduce the correlation among the training data, we proposed the linear whitening filter and the nonlinear quantizer.
Through the numerical simulations, we demonstrated significantly improved mis-classification performance.
These reveal the great potential of the proposed DMNNs to be adopted as a practical indoor localization technique.

\bibliographystyle{ieeetr}
\bibliography{Conference_mmWave_CS}
\end{document}